\documentclass[twocolumn]{aastex63}

\usepackage{amssymb}
\usepackage{natbib}

\newcommand{\kms}{{\rm km\,s}^{-1}}
\newcommand{\Msun} {M_\odot}

\newcommand{\jyb}{{\rm Jy\,beam}^{-1}}

\newcommand{\mum}{\rm {\mu}m}
\newcommand{\dotsec}{\rlap.{''}}

\received{June 1, 2019}
\revised{January 10, 2019}
\accepted{\today}
\submitjournal{APJ}

\shorttitle{Molecular contrail}
\shortauthors{Li \& Zhang}
\graphicspath{{./}{figures/}}

\begin{document}
\title{A mean density of $112\, M_{\odot}\,\rm pc^{-3}$  for Central Molecular Zone clumps -- Evidences for shear-enabled pressure equilibrium in the Galactic Center}

\correspondingauthor{Guang-Xing Li, Chuang-Peng Zhang}
\email{ gxli@ynu.edu.cn, cpzhang@nao.cas.cn}

\author{Guang-Xing Li}
\affiliation{South-Western Institute for Astronomy Research, Yunnan University,\\ Kunming, 650500 Yunnan, P.R. China}

\author{Chuan-Peng Zhang}
\affiliation{National Astronomical Observatories, Chinese Academy of Sciences, 100101 Beijing, P.R. China}
\affiliation{CAS Key Laboratory of FAST, National Astronomical Observatories, Chinese Academy of Sciences, 100101 Beijing, P.R. China}




\begin{abstract}
  We carry out a systematic study of the density structure of gas in the Central Molecular Zone (CMZ) in the Galactic center by extracting clumps from the APEX Telescope Large Area Survey of the Galaxy survey at 870\,$\mu$m.
  We find that the clumps follow a scaling of $m = \rho_0 r^3$ which corresponds to a characteristic density of $n_{\rm H_2}  = 1.6 \times 10^3\,\rm cm^{-3}$ ($\rho_0 =112\;M_{\odot}\;\rm pc^{-3}$) with a variation of $\approx 0.5\,\rm dex$,  where we assumed a gas-to-dust  {mass} ratio of 100. This characteristic density can be interpreted as the result of  {thermal} pressure equilibrium between the molecular gas and the warm ambient interstellar medium.   {Such an equilibrium  can  {plausibly} be established since shear has approximately the same strength as self-gravity.}  Our findings {may explain} the fact that star formation in the CMZ is highly inefficient compared to the rest of the Milky Way disk. We also identify a population of clumps whose densities are two orders of magnitudes higher in the vicinity of the Sgr\,B2 region, which we propose are produced by collisions between the clumps  {of lower densities}. For  {these} collisions to occur, processes such as compressive tides probably have created the appropriate condition by assembling the clumps together.
\end{abstract}



\section{Introduction}
The Central Molecular Zone (CMZ) is a disk-like gas structure that rotates
around the  center of the galaxy. The region has a size of $\approx  500\,\rm pc$,
and it contains a total of $3\times 10^7$\, {$\Msun$} of molecular gas
\citep{1987ApJS...65...13B,1998A&A...331..959D}. The gas rotates at a speed of
$\approx 200\,\rm\kms$ where the centrifugal force is likely to be balanced by
gravity from the central stellar bulge \citep{2013PASJ...65..118S}\footnote{The
  exact geometry of the region is still under debate
  \citep{2017MNRAS.465...45C}.}. Different from ``ordinary" molecular clouds \citep[e.g.][]{2015ARA&A..53..583H},  gas
in the CMZ is characterized by a higher degree of turbulent motion
\citep{2012MNRAS.425..720S}. It has been found that the star formation
of  {dense gas in the} CMZ is one order of magnitude lower
compared to that of the Milky Way
\citep{2013MNRAS.429..987L,2014MNRAS.440.3370K,2017MNRAS.469.2263B,2017A&A...603A..89K}.

Studying the evolution of gas in the CMZ is important for two reasons: first,
the fact that the gas dynamics and star formation in the CMZ are distinct from
the rest of the Milky Way means CMZ is a unique laboratory where we can
deepen our understanding of the star formation process.  Second, understanding
the evolution of gas in the CMZ is a key to understanding gas transport between the
Galactic disk and the central black hole,  {and can provide insights into
    other} important questions such as black hole growth, Active Galactic Nucleus (AGN) feedback, and galactic
disk evolution.

\begin{figure*}
  \centering
  \includegraphics[width=1.02  \textwidth, angle=0]{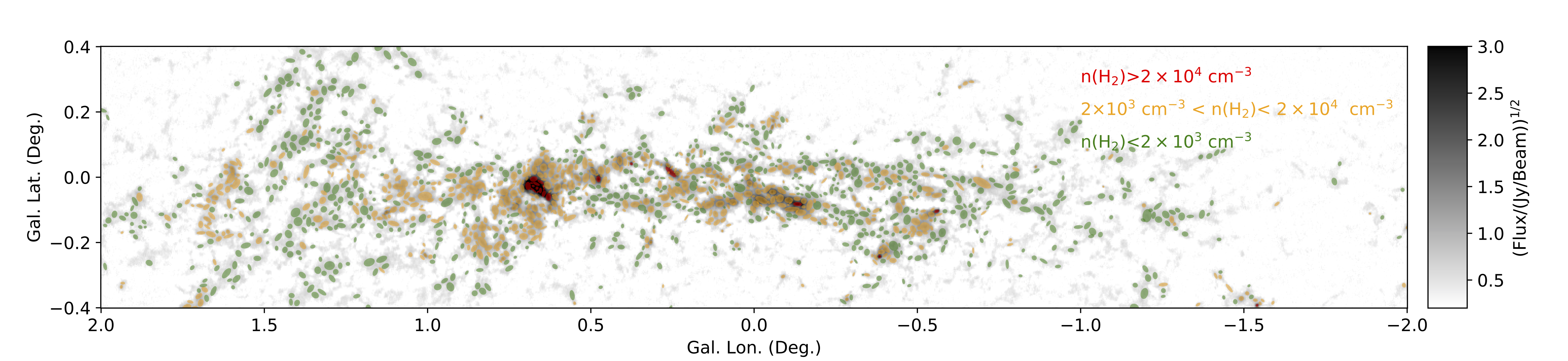}
  \caption{Spatial distribution of clumps of different densities. The grayscale image in the background is the 870\,$\mum$ continuum emission map from the ATLASGAL survey \citep{Schuller2009}.  Overlaid ellipses
    in different colors represent clumps of different densities extracted using the \texttt{GAUSSCLUMPS} algorithm.
    The densities are indicated at the upper right of the panel.}
  \label{Fig_GCLATLASGAL}
\end{figure*}

A very first step toward understanding the gas evolution is to  {study} the density structure, which  we focus on in this paper. There have been plenty of studies that characterize the (spatial, kinematic, and chemical) structure of molecular gas in the CMZ \citep[e.g.][]{2008MNRAS.386..117J,2010ApJ...721..137B,2015MNRAS.447.1059K,2016MNRAS.457.2675H}.
However, a detailed study on the density structure of the CMZ and its evolution down to sub-pc scale is still lacking.

In this paper, we study the density structure of the CMZ region using data from
the APEX Telescope Large Area Survey of the Galaxy
\citep[ATLASGAL;][]{Schuller2009}. ATLASGAL is a survey of the inner
Galaxy at 870\,$\mu$m performed by the APEX telescope
\citep{2006A&A...454L..13G}. It has a spatial resolution of around 19$''$.
  {Continuum observations at 870\,$\mu$m atmospheric window}  {are well-suited for}
tracing the cold gas. Besides, the 19$''$ resolution translates to a scale of
$0.8$\,pc assuming that we are $\approx 8.2\,\rm kpc$
\citep{2019A&A...625L..10G} away from the CMZ, guaranteeing that the majority of
the dense clumps at the CMZ region are reasonably resolved. Compared to previous
surveys such as the BGPS (Bolocam Galactic Plane Survey) \citep{2011ApJS..192....4A}, the ATLASGAL survey has a
much higher detection sensitivity ($\sigma=0.05 \,\jyb$). Taking advantage of
this, we perform a systematic study of the structure of gas in the whole CMZ
region, and perform a joint analysis of the statistical properties of the gas
clumps with their positions and discuss the implications.

\section{Observations and Analysis}
\label{sec:obs}

\subsection{Continuum data at 870\,$\mu$m}
\label{sec:data}

We use 870\,$\mu$m continuum map from ATLASGAL\footnote{The ATLASGAL project is
  a collaboration between the Max-Planck-Gesellschaft, the European Southern
  Observatory (ESO), and the Universidad de Chile.} survey
\citep{Schuller2009} to study the density structure of dense gas in the
CMZ.   The ATLASGAL data\footnote{ {Available at \url{https://atlasgal.mpifr-bonn.mpg.de/cgi-bin/ATLASGAL_DATABASE.cgi}.}} are well-suited to the study of small-scale (0.8$-$6
pc), high surface density  structures (structures whose surface densities are
larger than $140\;M_\odot \rm\,pc^{-2}$).  On the smaller side, we are limited by our resolution, which is
$ 0.8\;\rm pc$. On the larger side, structures of large sizes (larger than 2.$'$5, or  6\,pc) are
filtered out due to the limitation of ground-based bolometer observations \citep{Schuller2009}.
Structures with surface densities lower than $140\; M_\odot\; \rm\,pc^{-2}$ can
not be reliably detected due to our limited sensitivity ($\sigma=0.05 \,\jyb$).
  {Due to the limited spatial dynamical range,} we might underestimate the mass of the larges clumps
by up to 50\% \citep{2018arXiv180807499M}.

%

%
%

\subsection{Clump extraction}
To capture the density structure of the molecular gas and study its spatial
variations, we use the algorithm \texttt{GAUSSCLUMPS}
\citep{Stutzki1990,Kramer1998} in the
GILDAS\footnote{\url{https://www.iram.fr/IRAMFR/GILDAS/}} software package to
extract dense clumps. This method has
been successfully adopted in \citet{Zhang2018,Zhang2019}. We extract clumps by fitting Gaussians to a well-resolved peaks whose
intensities $I_{\rm870\mum}$  are above 5$\sigma$ ($\sigma=0.05 \,\jyb$) and whose FWHMs are
larger than 19$''$ (870\,$\mum$ beam size). The  1483 identified clumps are
plotted in Figure\,\ref{Fig_GCLATLASGAL}, and the physical
parameters are listed in Table\,\ref{tab_clumps}.

\begin{figure}
  \centering
  \includegraphics[width=0.49\textwidth, angle=0]{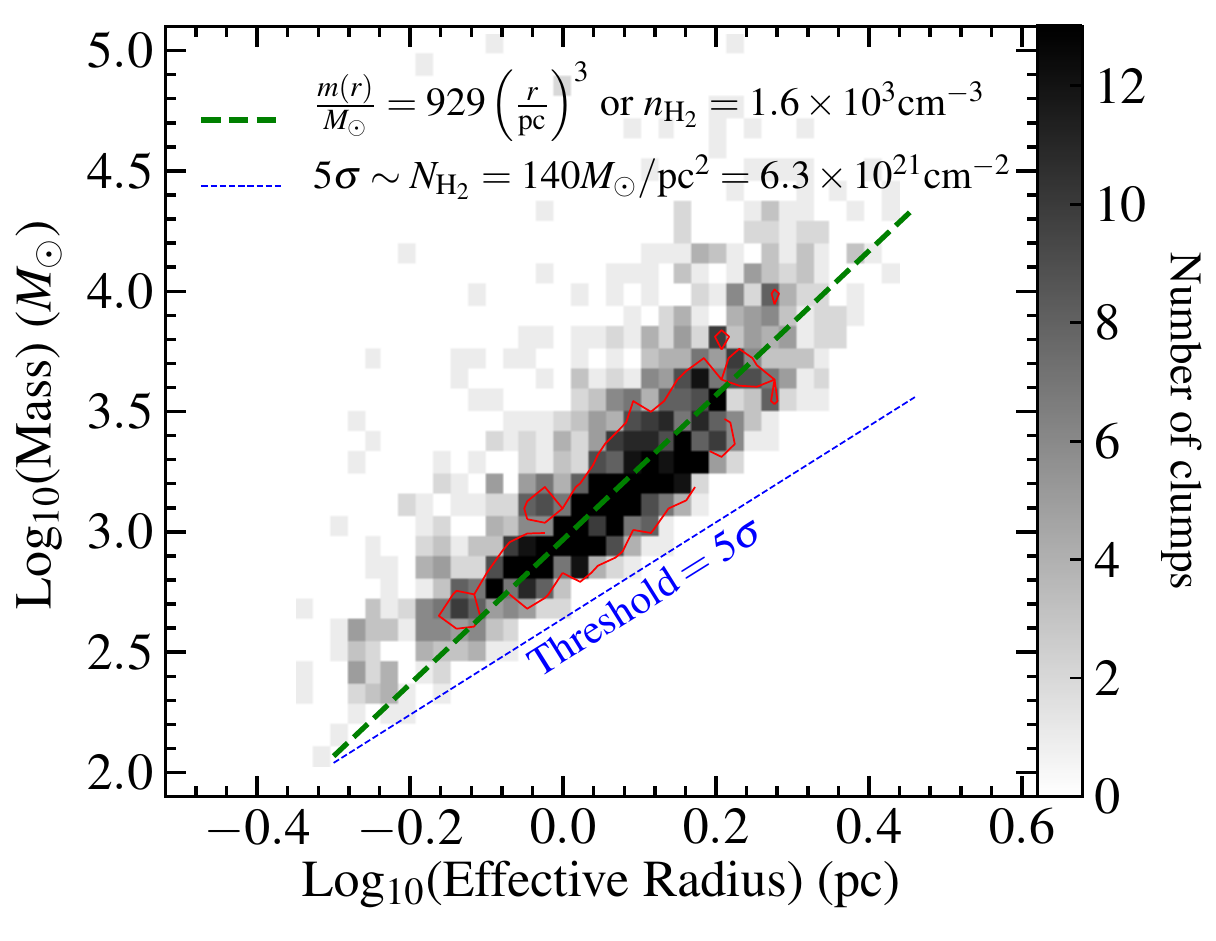}
  \caption{Mass$-$size distribution of the clumps with  a threshold $I_{\rm870\mum} >5\sigma$ (indicated by blue dotted line) in the CMZ region. The green dashed line stands for $m_{\rm clump}/M_{\odot} = 929\, (r/{\rm  pc})^3$, which corresponds to $n_{\rm H_2}= 1.6\times 10^3\rm\,cm^{-3}$.  {Note that to compute the volume density, we have used Eq. \ref{eq:vol}, whose justification can be found in Appendix \ref{estimate:rho}.}
  The red contour represents locations with more than seven clumps per pixel.  See Sec. \ref{sect:mass} for details.}
  \label{Fig_mass_size}
\end{figure}

\begin{figure}
  \centering
  \includegraphics[width=0.49\textwidth, angle=0]{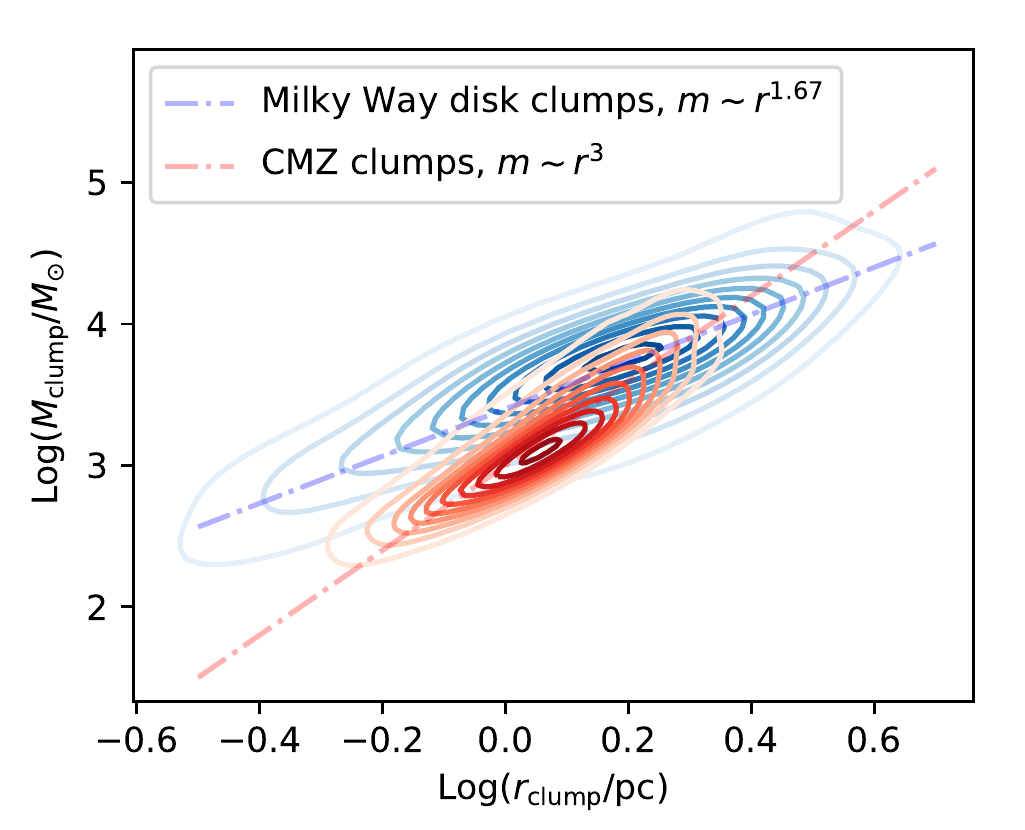}
  \caption{ {Clumps$-$scale mass$-$size relation. The blue contours represent the distribution of proto$-$cluster clumps in the mass-size plane. The data come from \citep{Urquhart2013}. The blue line represents the scaling $m_{\rm clump} \propto r_{\rm clump}^{1.67}$ proposed in \citep{Li2017}. The red contours represent the distribution of CMZ clumps in the mass-size plane, and the red line represents the scaling $m_{\rm clump} \propto r_{\rm clump}^{3}$ found in this work. The contours are constructed using density fields estimated from the KDE (kernel density estimation) algorithm.} \label{fig:mrcompare}}
  \label{Fig_mass_size_compare}
\end{figure}

\subsection{Dust temperature estimation}
\label{sect:temperature} To accurately derive the clump mass it is necessary
to estimate the dust temperature. Using high-quality Hi$-$GAL (Herschel infrared Galactic Plane Survey) data covering a
large wavelength ranging from 70 to 500\,$\mum$ \citep{Molinari2016},\footnote{ {Available at \url{https://irsa.ipac.caltech.edu/applications/Herschel/}.}} we
calculate the dust temperature map via fitting the spectral energy distribution
(SED) extracted from the multiwavelength images on a pixel-by-pixel basis
\citep[e.g.][]{Wang2015}. This method has been successfully adopted in many
works, such as \citet{Zhang2017irdc,Zhang2017g18} and \citet{Zhou2019}.  {
In our calculations, a smooth component is removed using a Fourier
transform-based approach \citep[][]{Wang2015}. In this method, the original
images have been transformed into the Fourier domain and separated into the low
and high spatial frequency components, and then inversely transferred
back into the image domain. The low-frequency component corresponds to large-scale
background/foreground emission, while the high-frequency component represents
the emission of interest.  In spite of this, we might still be contaminated by
clumps in the Milky Way disks. However, these clumps are small in number
\citep[10 \%, estimated from][]{2014A&A...565A..75C} and do not contribute significantly to
our statistics.} In current work, we regridded the maps to the same resolution of
11$\dotsec$5, and convolved the images with the same Gaussian beam with $\mathrm{FWHM
    = 45''}$, corresponding to the measured beam size of 500\,$\mum$ data. Other
parameter setup is the same as that in \citet{Zhang2017irdc}. The dust
temperature $T_{\rm dust}$ map was shown in Appendix \,\ref{sect:a1}.
\subsection{Mass and density calculation}
\label{sect:mass}
Assuming that the dust emission is optically thin,
we calculate the clump masses  $M_{\rm H_2}$ following \citet{Kauffmann2008} via
\begin{eqnarray}
  \label{equa_mass}
  \left(\frac{M_{\rm H_2}}{\Msun}\right)   = & 0.12 \left({\rm  e}^{14.39\left(\frac{\lambda}{\rm mm}\right)^{-1}\left(\frac{T_{\rm dust}}{\rm K}\right)^{-1}}-1\right) \times &  \nonumber \\
  & \left(\frac{\kappa_{\nu}}{\rm cm^{2}g^{-1}}\right)^{-1} \left(\frac{S_{\nu}}{\rm Jy}\right) \left(\frac{D}{\rm kpc}\right)^{2}
  \left(\frac{\lambda}{\rm mm}\right)^{3} \left( \frac{\eta_{\rm gas}}{100} \right) \;,
\end{eqnarray}
where $\lambda=870\,\mum$ is the observational wavelength, $T_{\rm dust}$ is the dust temperature (see Section\,\ref{sect:temperature}), $\kappa_{\nu} = 0.0185 \,\rm cm^2 g^{-1}$ is the dust opacity at 870\,$\mum$ \citep{Ossenkopf1994}, $D$ is the distance to the Sun, $\eta_{\rm gas}$ is the gas-to-dust ratio  and integrated flux $S_{\nu}$ is
\begin{equation}
  S_{\nu} = I_{\rm peak} \times \frac{\rm FWHM_{ext}^2}{\rm FWHM_{\rm obs}^2},
\end{equation}
where ${\rm FWHM_{ext}}$ is the extracted Gaussian size of each clump by \texttt{GAUSSCLUMPS}, and ${\rm FWHM_{obs}}$ is the beam size of the ATLASGAL observations.  We assume that all the clumps we analyze sit at a distance of 8.2 kpc \citep{2019A&A...625L..10G}.

The surface density of the clumps are estimated as
\begin{equation}\label{eq:sigma}
  \Sigma_{\rm clump} = m_{\rm clump} / \pi r_{\rm clump}^2\;,
\end{equation}
where $r_{\rm eff}$ is
related to FWHM by $r_{\rm eff} = {\rm FWHM}/(2 \sqrt{{\rm ln2}})$. The $\rm H_2$ column density is related to the surface density by
  \begin{equation}
    N({\rm H}_2) = \Sigma_{\rm H_2} / (\mu_{\rm H_2}
    m_{\rm H})\;,
  \end{equation}
  where $\mu_{\rm H_2} \approx 2.8$ is the
  mean molecular weight \citep[e.g.][]{Kauffmann2008}.

  If the cores are considered as uniform spheres, the volume density $n_{\rm
    H_2}$ can be estimated by

  \begin{equation} \label{eq:vol}
    \rho_{\rm clump} = 0.12 \;
    m_{\rm clump}/{r_{\rm eff}^3}\;,
  \end{equation}
  a
  detailed justification of this formula can be found in Appendix
  \ref{estimate:rho}.
  {For convenience, we  also express the  gas density in
  terms of the number of $\rm H_2$ molecules found per cubic centimeter,
  where}
  \begin{equation} n_{\rm H_2 } = \frac{\rho_{\rm clump}}{\mu_{\rm H_2}
      m_{\rm H}}\;,
  \end{equation}
  {  The derived parameters
  are listed in Table\,\ref{tab_clumps}.

  In our fiducial calculations, we make the working assumption that the gas-to-dust mass ratio $\eta_{\rm gas}$ is 100, which is essentially the value reported in the solar neighborhood. We have adopted this assumption to ensure that the values we derived are comparable to the values reported from the other authors. Nevertheless, this assumption is unlikely to be accurate, as the gas$-$to$-$dust ratio is expected to evolve significantly with respect to metallicity and hence Galactocentric distance. By extrapolating results from  \citet{2017A&A...606L..12G}, the gas-to-dust ratio in the Galactic center might be lower than 50. This uncertainty must be kept in mind when interpreting the results.

  \begin{figure*}
    \centering
    \includegraphics[height=0.285\textwidth, angle=0]{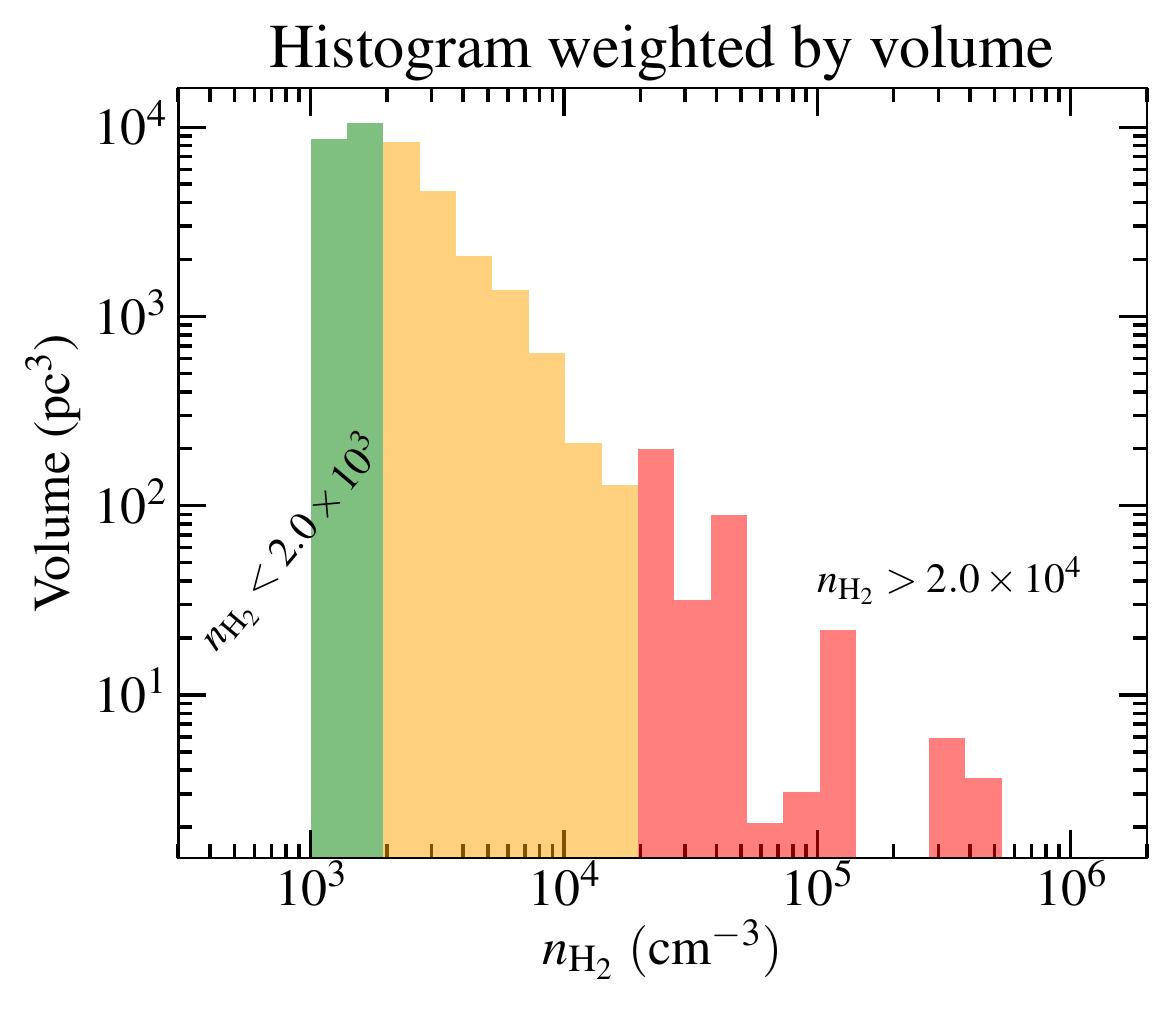}
    \includegraphics[height=0.285\textwidth, angle=0]{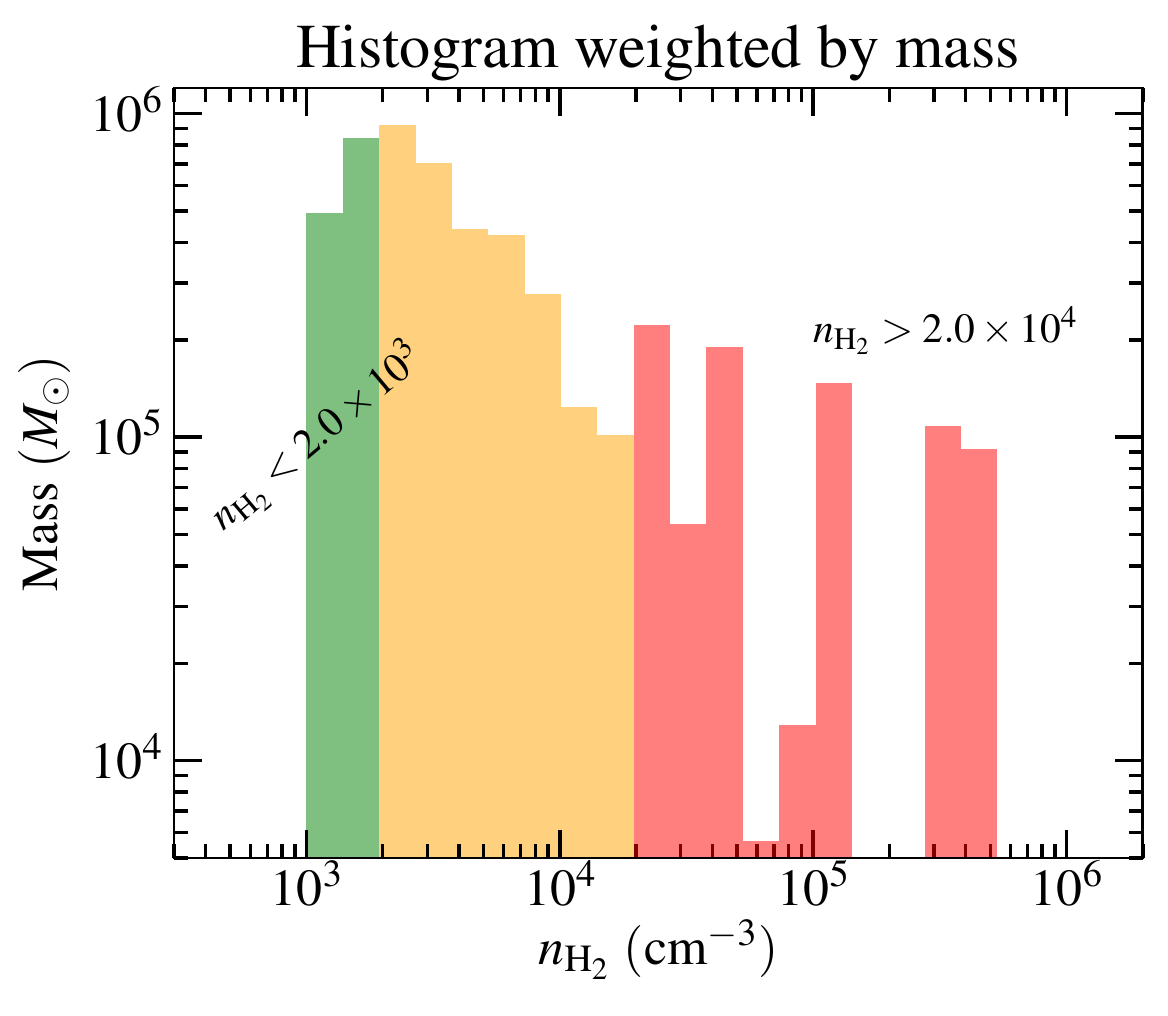}
    \includegraphics[height=0.285\textwidth, angle=0]{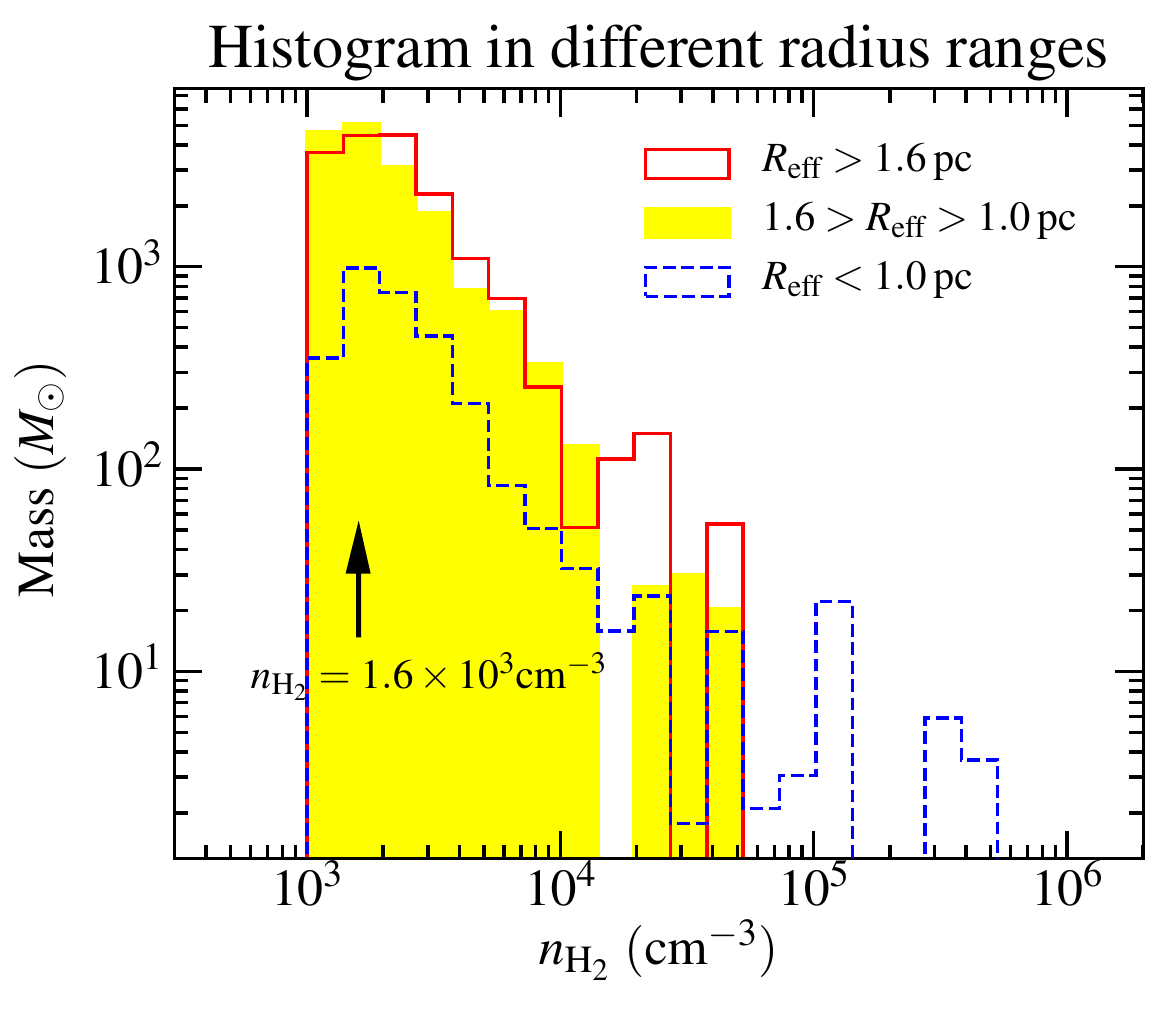}
    \caption{Density distribution of the CMZ region. Left panel: Volume-weighted density histogram of the CMZ region. Middle panel: Mass-weighted density histogram of the CMZ region. In these panels,   {contributions from }clumps of different densities are indicated with different colors. Right panel: Volume-weighted density histogram of clumps of different sizes. Clumps of different sizes seem to share a common mean density of $n_{\rm H_2} = 1.6 \times 10^3\,\rm cm^{-3}$.}
    \label{Fig_hist_density}
  \end{figure*}

  \section{Results}

  \subsection{Mass, Size and density distributions}
  \label{sect:mass_size_density}  We study the density distribution of gas on scales from $0.8$ to $6$ pc and column density $\gtrsim 140\; M_{\odot}\; {\rm pc}^{-2}$, which can be reliably recovered by the ATLASGAL survey.
  Figure\,\ref{Fig_GCLATLASGAL} plots the results of our clump extraction
  obtained using the \texttt{GAUSSCLUMPS}, where each clump is represented with an
  ellipse. Our source extraction has captured the majority significant
  structures ($I_{\rm870\mum}>5\sigma$) visible on the map. In total, we have
  recovered a mass of $5.38\pm0.05\times10^6\,\Msun$, which is much larger than
  the mass of $6 \times 10^5\,\Msun$ recovered by the BGPS survey
  \citep{2010ApJ...721..137B}.
  However, our total
  mass is still smaller than that ($\approx 3 \times 10^7\,\rm \Msun$) reported in
  \citet{1987ApJS...65...13B} and \citet{1998A&A...331..959D} estimated using CO
  observations of a much lower resolution. This difference is caused by a
  combination of: (1) they analyzed an area that is twice as large as ours, and (2)
  although  our observations have significantly better angular resolutions, we
  are more limited in sensitivity and spatial dynamical range. The $^{13}$CO
  observations of \citet{1987ApJS...65...13B} allow them to probe gas with much
  lower surface densities (e.g. a few tens of $M_\odot \rm\,pc^{-2}$), which is
  around 1 order of magnitude lower than our limiting surface density.
  Presumably, there are gases whose surface densities lie between 50 to
  140\,$M_\odot \rm\,pc^{-2}$. These gases can be detected (yet unresolved) in
  \citet{1987ApJS...65...13B} but remain undetected by ATLASGAL.

  %
  Figure\,\ref{Fig_mass_size} plots the distribution of clumps in the mass-size
  plane. This distribution is clearly different from that of the ``ordianry'' ATLASGAL clumps (Figure\,\ref{Fig_mass_size_compare}). 
    Above the detection limit, the clumps have a  structured
  distribution, where pixels that contain the largest number of
  clumps seem to follow a relation with $m_{\rm clump} \propto r_{\rm clump}^3$, which points to a constant density that is independent on the scale. 
    We note that this characteristic density is somehow dependent on whether we weight the density distribution by volume or by mass.

  %

  To further study the density structure of the region, in
  Figure\,\ref{Fig_hist_density}, we plot both the volume-weighted and
  mass-weighted density distribution of the clumps. From the volume-weighted plot,
  we measure a characteristic density of  $n_{\rm H_2} = 1.6 \times
10^3\rm\,cm^{-3}$, $\rho_0 = 112\,M_{\odot}\rm \,pc^{-3}$  by fitting a
  log-normal function to the distribution and a density variation of $\approx  0.5$
  dex is measured from the FWHM of the distribution. This corresponds to a mass-size relation of
  \begin{equation}
    \label{eq:mr}
    m_{\rm clump}/M_{\odot} = 929 (\eta_{\rm gas}/100) (r/{\rm pc})^3  \, ,
  \end{equation}
  {where the density and the normalization of the mass-size relation is related by Eq. \ref{eq:vol}}.

  To  further demonstrate the
  existence of this characteristic density, in the right panel of Fig.
  \ref{Fig_hist_density}, we plot the mass-weighted density distribution of
  groups of clumps of different sizes. Since the peaks of the density distributions from these subsamples do not evolve with the clump radii, the characteristic density that we derived is an intrinsic property shared by these clumps.

  \subsection{Spatial distribution of clumps of different densities}

    Although our clumps sample exhibits a density variance of $\approx 0.5$ dex, a
        significant number of clumps have densities that are two orders of
        magnitudes larger than the mean density. To better understand the density
        structure of the region, in Figure\,\ref{Fig_hist_density}, we divide the
    clumps into three groups: the first group of clumps has a density of
    $n_{\rm H_2} < 2.0\times10^3 \,\rm cm^{-3}$, which is called ``lower-density
    clumps", the second group of clumps has a density of $2.0\times10^4 >
      n_{\rm H_2} > 2.0\times10^3 \,\rm cm^{-3}$, which we call ``higher-density
    clumps", and the third group of clumps has a density of $n_{\rm H_2}
      \gtrsim 2.0\times10^4 \,\rm cm^{-3}$, which is called as ``highest-density
    clumps". The ``highest-density clumps" seems to belong to a parameter range
    that is separated from the majority of the clumps indicated by the
    discontinuity of density distribution in Figure\,\ref{Fig_hist_density}.

  We then plot the spatial distributions of clumps of different densities in
  Figure\,\ref{Fig_GCLATLASGAL}.  It emerges that the spatial distribution of
  the higher-density clumps exhibits a pattern where they seem to follow an
  arc-like structure which stretches from $l=1^{\circ}$ to $l=-1^{\circ}$, and
  it contains some of the most active star-forming regions in the CMZ like the
  Sgr\,B2. To some extent, one can relate our dense arc to the 100-pc twisted
  ring identified by \citet{2011ApJ...735L..33M}, where dense gas (gas with
  densities above $n_{\rm H_2} \approx 2.0 \times 10^3 \,\rm cm^{-3}$) forms a
  coherent, twisted pattern, and we note that  {in addition to that,} gas is unevenly distributed along
  this ring, with clumps of the highest densities  ($n_{\rm H_2} \gtrsim
2.0 \times 10^4 \,\rm cm^{-3}$)  distribute  mostly within the vicinity of the
  Sgr\,B2 region. { This uneven distribution of dense gas implies that the
  dynamics of the region are non-stationary, as expected from some recent models
  \citep[e.g.][]{2015MNRAS.447.1059K,2016MNRAS.457.2675H,2018MNRAS.475.2383S}.

  \section{Discussions}
  \subsection{ {Mass-size relation as density structure diagnostics}}
  We focus on the mass-size relation of clump-scale dense gas revealed by the ATLASGAL survey. Our mass-size relation should be distinguished from the core-scale mass-size relation seen in \citep[e.g.][]{2008ApJ...672..410L}, and various cloud-scale mass-size relation \citep[e.g.][]{2010ApJ...723..492R} summarized in \citet[][]{2020MNRAS.tmp..225C}. We remind the reader that we use ATLASGAL observations at 870$\mu$m
  obtained with the APEX telescope. Due to the fact that the emission from dust
  is optically thin, the ATLASGAL survey data is ideal for tracing the
  distribution of dense gas in the Milky Way
  (where $\Sigma_{\rm gas}\geq 140\, M_{\odot}\;\rm pc^{-2}$).
  In the Milky Way disk, all the dense clumps seen
  in ATLASGAL should  {collapse to form star clusters in a few crossing time \citep{2012A&A...542L..15W,2016A&A...585A.149W,Li2018} whereas in the Milky Way center they do not appear to be collapsing.}

  On the clump scale, structures in the Milky Way disk follow
$m \propto r^{1.67}$
  \citep[e.g.][]{Urquhart2013,Pfalzner2016,Li2017},
  meaning that larger clumps have  smaller densities.
  In contrast to this, in  the Galactic center, the clumps seem to have a
  density that is independent on the clump size. This reflects the uniqueness of
  the CMZ in terms of density structure. A comparison is made in  Fig. \ref{fig:mrcompare}.

  %

  \subsubsection{Evidence of  pressure equilibrium}
  We propose that this almost-constant gas density observed in the CMZ region can be explained by the  {thermal} pressure equilibrium. The gas temperature is found to be around $50 - 100\,\rm K$ \citep{2007PASJ...59...25N,2013A&A...550A.135A,2016A&A...586A..50G}. Assuming a temperature of $70\,\rm K$, we estimate that the cold gas has a pressure of
  \begin{equation}
    \frac{p_{\rm internal}} {k_{\rm B}}= n_{\rm H_2} T_{\rm gas} (\eta_{\rm gas} / 100)\approx 1.1 \times 10^5\,\rm K\,cm^{-3}\,,
  \end{equation}
  where we assumed $n_{\rm H_2} = 1.6 \times 10^3\;\rm cm^{-3}$,
$k_{\rm B}$ is the Boltzmann constant.

  As a comparison, using temperature measured by
  \citet{1990ApJ...365..532Y}, \citet{1992Natur.357..665S} estimated a
  pressure of
  \begin{equation} p_{\rm external} / k_{\rm B} \approx {10^5}\textup{--}
    {10^6} \,\rm K\,cm^{-3}\,, \end{equation}
  for the warm ambient gas,
  where the pressure is computed using $p = n k_{\rm B} T$. The
  temperature and density can be estimated by modeling the X-ray emission, and the  {pressure we estimated is broadly consistent} with results reported in the literature
  \citep[e.g.][]{2004ApJ...613..326M,2015MNRAS.453..172P,2019Natur.567..347P,2019ApJ...875...32N}.
  Since $p_{\rm internal} \approx p_{\rm external}$,  {thermal} pressure equilibrium
  does provide a good explanation to the observed gas density.

  We still observe a density fluctuation of around $ 0.5\,\rm dex$,
  which can be caused by processes such as turbulence. Turbulence is known
  to be capable of producing density variations
  \citep{1994ApJ...423..681V,1997ApJ...474..730P,1998ApJ...504..835S,2010A&A...512A..81F}.
  The clumps we observed have a (volume-weighted) density variation of
$\approx 0.5 \,\rm dex$. Adopting the relation between density variation
  and Mach number from \citet{2012ApJ...761..149K}, this can be produced by
  turbulence with a Mach number of around  {2}. This requires a turbulent
  velocity dispersion of a few $\rm km\, s^{-1}$  {for solenoidal turbulence}, where  we adopt an isothermal sound speed of around $0.5\;\rm
    km\, s^{-1}$ (the sound speed is derived assuming a temperature of 70 K). This is
    {smaller yet comparable to the} typical clump-scale velocity dispersion found in
  \citet{2012MNRAS.425..720S} measured at a scale of around $ 1\;\rm pc$.  Turbulence is capable of producing the
  observed density variations.

  Finally, we note that in our picture, an equilibrium is established
  between the thermal pressure of cold gas in the clump and the thermal
  pressure of the ambient hot gas, and turbulence   {can only} create
  additional density variations. Our view is shared by  \citet[][]{2003ApJ...587..278W}.

  \subsubsection{Regulation of collapse by shear}
  Even in the Milky Way disk,  {thermal} pressure equilibrium does play an important role in setting the density of the molecular gas \citep{1968ApJ...152..971S,1969ApJ...155L.149F},  {however, clumps in the Galactic disks are dense objects that sit at the centers of molecular clouds, and their densities are mainly determined by an interplay between turbulence and gravity \citep{Li2017}.  In contrast, the clumps in the CMZ share a density that is mostly independent of the clump radius. We propose that this can be explained in a {\it shear-enabled pressure equilibrium} scenario where  shear is able to counteract against gravitational collapse.  }

  In the absence of shear, self-gravity is dynamically important for these clumps. We first estimate the  {effective} pressure caused by gravity, which is
  \begin{eqnarray}
    \frac{p_{\rm gravity}}{k_{\rm B}} &\approx&  \frac{G m_{\rm clump}^2} { k_{\rm B}   r^2  } \times \frac{1}{ 4 \pi r^2} \approx \frac{G \pi \Sigma_{\rm clump}^2} {4\,k_{\rm B}}  \approx \frac{ 9  G \pi \rho^2 r_{\rm clump}^2}{4\,k_{\rm B}} \nonumber \\
    &\approx& 2 \times 10^6 \, {\rm K\,cm^{-3}} \Big{(} \frac{r_{\rm clump}}{\rm pc}\Big{)}^2 \Big{(}\frac{\eta_{\rm gas}} {100}\Big{)}^2 \,,
  \end{eqnarray}
  where $r_{\rm clump}$ is the size of a clump,  {and we have assumed $\Sigma_{\rm clump} = m_{\rm clump} / (\pi r_{\rm clump}^2)$, $\Sigma_{\rm clump} \approx 3 \rho_{\rm clump} r_{\rm clump}$\footnote{ {Derived by combining Eq. \ref{eq:sigma} and Eq. \ref{eq:vol}.}}
      ,  and $\rho_{\rm clump}= 112 M_{\odot}\;\rm pc^{-3}$}. For a typical clump of size $\approx 1\,\rm pc$, this pressure is indeed comparable to the internal and external pressure we estimated before,  {and one would naively expect gravity to be able to compress the  gas significantly.}

  Although gravity should be  {a major player}, in the CMZ, its effect is largely canceled by effects like shear and extensive tidal force. Shear can cause gas at  different radii to rotate at different angular speeds, and this differential motion stretches gas into long streams before they can collapse on their own. Tidal force causes different parts of a clump to accelerate differently, and it can halt the fragmentation when the tidal force is extensive.

  The fact that gas clumps organize into streams \citep[e.g.][]{2016MNRAS.457.2675H} indicates that shear should be dynamically important. To evaluate the relative importance between shear and self-gravity, we compute the  {relevant timescales}. Assuming that the rotation curve of the CMZ gas can be parameterized as $v\propto r^{p}$, the shear time is

  \begin{eqnarray} \label{eq:shear1}
    t_{\rm shear} &=&
    \Big{(}\frac{\partial \Omega}{\partial r} r \Big{)} ^{-1} =   |\frac{1}{1-p}|\, \Omega^{-1}  \nonumber \\
    &\approx& 1.1 \times  \Omega^{-1}
    \approx \; 1.7\, |1-p|^{-1}  \,\rm Myr \nonumber \\
    &\approx&  2.8\;\rm Myr\,,
  \end{eqnarray}
  where  we have assumed $p\approx 0.4$. To estimate the value of $p$, we use the mass profile derived by \citet{2015MNRAS.447.1059K}, which is based on the mass profile of \citet{2002A&A...384..112L}.

  For comparison, the time for gravitational collapse to occur on a sphere of a constant density is  

  \begin{eqnarray}\label{eq:ff}
    t_{\rm ff} &=& \sqrt{\frac{3\pi}{32 G \rho_{\rm gas}}} \nonumber \\
    & \approx & 0.54 \times \sqrt{\frac{1}{G\rho_{\rm gas }}} \approx 0.7 (\eta_{\rm gas}/100)^{-1/2} \,\rm Myr
  \end{eqnarray}
  {where we have used $\rho_{\rm gas } = 112 M_{\odot}\;\rm pc^{-2}$},
  where $\eta_{\rm gas}$ is the gas-to-dust mass ratio, and its value is still
  uncertain and can range from 10 to 100. To interpret these numbers, we note that
  both the shear time and the freefall time are subject to significant
  uncertainties. The shear time is sensitive to the shape of the rotation profile
  measured in terms of $p$ in Eq. \ref{eq:shear1}.The version of
  the rotation profile used in this paper is derived using the stellar mass model presented in
  \citet{2002A&A...384..112L} and a discussion on its uncertainty seems to be
  missing. The freefall time, which is dependent on the gas density and hence the gas-to-dust ratio, is also uncertain (Sec. \ref{sect:mass}). Assuming that $p=1$ and
$\eta_{\rm gas} = 100$, the shear time is around 3 Myr, which appears to be $\approx 4$
  times the freefall time. However, since recent observations seem to indicate
  that the  $\eta_{\rm gas}$ can be as low as $10$ in the Galactic center, it
  seems plausible that the shear time is comparable to the freefall time. The
  current data is consistent with the proposal that shear can counteract
  against gravitational collapse in the Galactic center region.


  Our findings provide crucial insights into the puzzle of the inefficiency of
  star formation in the CMZ region. It is believed that enhanced
  turbulence is responsible for the inefficiency of star formation
  \citep{2014MNRAS.440.3370K}. Here, our analyses have revealed that in the CMZ
  region, shear has the adequate strength to halt the collapse of the individual
  clumps. Our results agree with earlier proposals that shear can be a factor to
  regulate collapse  {in the CMZ region}
  \citep[e.g.][]{2002A&A...384..112L,2013MNRAS.429..987L,2015MNRAS.446.2468E,2015MNRAS.453..739K,2018MNRAS.478.3380J},
  although in those papers shear is expected to  halt the collapse of the CMZ on
  the large scale,  {preventing it to collapse globally}, whereas in our case,  the importance of shear is more
  pronounced where it can stop the collapse of the individual clumps, although
  other processes such as turbulence can also be important.


  \subsection{Forming dense gas through clump collisions}

  Although the densest clumps
  ($n_{\rm H_2} \gtrsim 2.0\times10^4 \,\rm cm^{-3}$) only contains a small fraction of the mass, they are
  associated with the majority of star formation found in this region
  \citep[see, e.g., the star formation rate of individual clouds
    from][]{2017A&A...603A..90K}.  Interestingly, they are
  distributed almost exclusively in the vicinity of the Sgr\,B2 region
  (Figure\,\ref{Fig_GCLATLASGAL}).




  We propose that a preferred  way to produce these dense clumps is through clump
  collisions. It is believed that collision between molecular clouds should occur regularly in
  the Milky Way disk
  \citep{2009ApJ...700..358T,2011ApJ...738...46T,2014ApJ...780...36F,2017ApJ...835L..14G}.
  \citet{2011MNRAS.413.2935D} and \citet{2017MNRAS.471.2002L} estimated the
  timescale for such processes to occur and found that the collision time is
  comparable to the dynamical time in the bulk of the Milky Way disk. In the CMZ
  region, the collisions can produce the dense clumps since (a) clump collisions
  should occur in this region regularly, and (b) these collisions are capable
  of producing clumps of such high densities.


  To evaluate whether cloud$-$cloud collisions should occur near Sgr\,B2, {the mean free path can be estimated as  \citep{2017MNRAS.471.2002L}
      \begin{equation}
        \lambda_{\rm clump} \approx \frac{\Sigma_{\rm clump}}{\rho_{\rm mean}} = \frac{100 \, M_{\rm \odot}\,\rm pc^{-2}}{1000 \, { M_{\odot}} {\rm pc}^{-3}} \approx 0.1 \,\rm pc\,,
      \end{equation}
      where,} to estimate the mean surface density of the clumps, we have adopted a mean density of $n_{\rm H_2} = 2\times 10^3 \,\rm cm^{-3}$ and a typical size of $1\,\rm pc$ { (see Figure\,\ref{Fig_GCLATLASGAL})}, such that $\Sigma_{\rm clump} = 100\, {M_{\odot}}\,\rm pc^{-2}$ and the Sgr\,B2 region is estimated to have a mean density of
$\approx 1000\, M_{\odot}\,\rm pc^{-3}$ \citep[e.g.][]{2016A&A...588A.143S}.
  The mean free path of clumps in the region is around 0.12 $\,\rm pc$, which is much smaller than the size of the region. { If the clumps in the CMZ are not collapsing by themselves, when assembled to a very small region, collisions must occur. }

  Can clump collisions produce these highest-density clumps? In the
  vicinity of the Sgr\,B2, the density enhancement due to collisions can be
  estimated as $\rho ' /\rho \approx \mathcal{M}^2 = (v_{\rm collide} / c_{\rm
    s})^2$ where $\rho ' /\rho$ is the density contrast. To produce a density
  enhancement of 100, one needs the clumps to collide at Mach number $\mathcal{M}=
10$. Assuming a sound speed of $0.5\,\rm \kms$, the cloud must collide with a
  relative speed of a few km/s, which is possible since an inter-clump
  velocity dispersion of a few tens of $\rm \kms$ is common at the CMZ region
  \citep{2016MNRAS.457.2675H}. Our mechanism of producing
  dense clumps through
  collisions is consistent with the observtional result of \citet{2015PASJ...67...90T} where
  they found enhancements of SiO emission lines in the vicinity of the Sgr\,B2
  complex, which they interpreted as the result of shocks produced during
  collisions.

  \begin{figure} \includegraphics[width=0.51\textwidth]{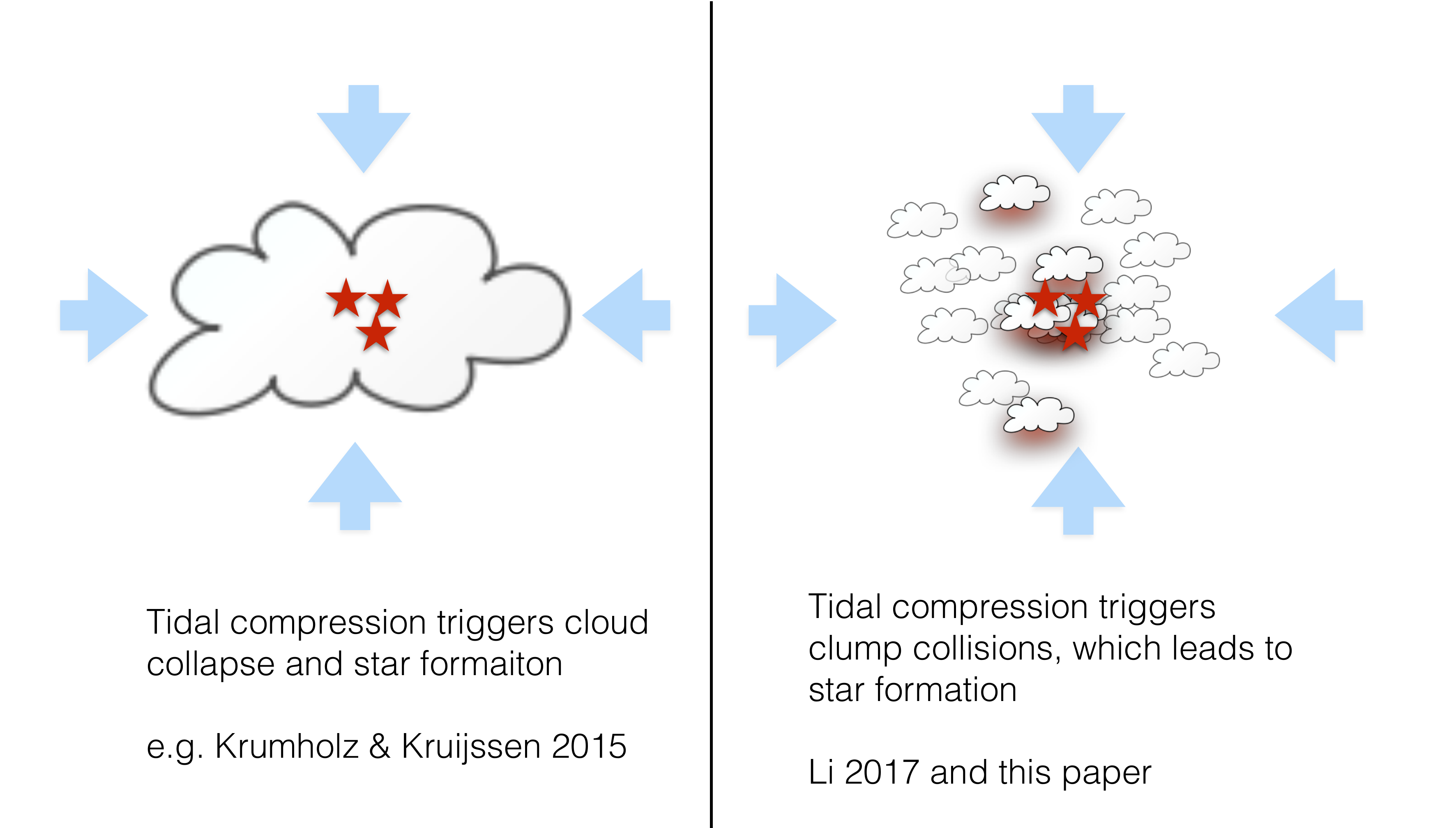}
    \caption{
      Different views on the origin of dense gas near Sgr B2. In
      previous models
      \citep{2015MNRAS.453..739K,2018MNRAS.478.3380J,2019MNRAS.486.3307D,2019MNRAS.484.5734K},
      it is believed that tidal force can compress gas clouds which leads the
      formation of dense gas and stars. In this paper
      \citep[following][]{2017MNRAS.471.2002L}, a tidal force can compress an
      ensemble of clumps,  {which triggers}  collisions that lead to the formation of clumps of much higher
      densities.} \label{fig:regimes} \end{figure}

  Recent papers have pointed out that the importance of processes such as changes
  of shear in triggering star formation
  \citep{2015MNRAS.453..739K,2018MNRAS.478.3380J,2019MNRAS.486.3307D,2019MNRAS.484.5734K}.
  However,  the
  picture we are proposing is very different.  In models such as
  \citet{2019MNRAS.486.3307D}, when a cloud passes through certain locations,
  tidal compression is imposed on the cloud as a whole which causes it to
  collapse, whereas in our case \citep[similar to][]{2017MNRAS.471.2002L}, an
  external compression cause ensembles of clumps  to collide and agglomerate, through
  which dense gas is produced and star formation is triggered. The collisions also
  lead to the formation of clumps of higher masses. The different scenarios are illustrated in Figure\,\ref{fig:regimes}.

  \section{Conclusion}
  Using data from the ATLASGAL survey, we study the density structure of the molecular gas in the CMZ region traced by dust continuum emission. We have extracted 1483 clumps from the data, and have studied the properties of the clumps in terms of mass, size, and density. We find that the majority of the clumps follow $m \propto r^3$, which points to a constant density of $n_{\rm H_2} = 1.6 \times 10^3\,\rm cm^{-3}$ ($\rho_0 = 112\;\rm M_{\odot}\,pc^{-3}$) {where we have assumed a dust-to-gas ratio of 100}. Clumps in localized regions such as the Sgr\,B2 vicinity have densities that are two orders of magnitudes higher than the mean density of gas in the CMZ.

  We  {propose} that this characteristic density can be explained by a {\it shear-enabled pressure equilibrium} model where the density is set by  balance between the thermal pressure cold gas and that of the warm ambient medium. Different from ``ordinary'' clouds in the Milky Way disk, in the CMZ, a  {thermal} pressure equilibrium can be achieved since shear (and possibly extensive tidal force) caused by gravity from the Galactic Bulge is strong enough to counteract against self-gravity.

  Our shear-enabled pressure equilibrium scenario can explain the inefficiency of star formation of the CMZ, as shear has the adequate strength to counteract against gravitational collapse,  {and this mechanism is expected to reduce the star formation significantly}. {Although other regulating mechanisms, such as  turbulence might still be playing important roles, the fact that the clumps in the CMZ follow a relation with $m_{\rm clump}\propto r_{\rm clump}^3$ seems to indicate that the role of shear is indispensable.

  We also identified an over-abundance of clumps with $n_{\rm H_2}> 2.0 \times 10^4\,\rm cm^{-3}$ in the vicinity of the Sgr\,B2 region. We propose that they are produced by agglomerations/collisions of clumps of lower densities. For collisions to occur, processes such as  tidal compression have probably provided the appropriate condition by assembling the clumps together.

Our analyses reveal that the gas in the CMZ belongs to a unique regime where shear is  {sufficient} to overcome gravity   in the individual clumps, such that the density of these clumps is determined directly by the  {thermal} pressure equilibrium  {established with respect to} the ambient environment. Our picture is crucial for understanding the evolution of gas at centers of other galaxies where star formation appears to be  {suppressed}.

\section*{Acknowledgements}
This work is partly supported by the National Natural Science Foundation of China 11703040. We thank Dr. Jinghua Yuan for the data reduction of dust temperature. G.-X. Li thanks Doug. N. C. Lin and Eugene Churazov for stimulating discussions.
We thank our referee whose detailed comments help to improve  our paper significantly.

\bibliographystyle{apj}
\bibliography{paper}

\appendix
\section{Temperature map} \label{sect:a1}
In Figure\,\ref{Fig_Tdust}  we present the map of dust temperature. See Section\,\ref{sec:obs} for details.

\begin{figure*}
  \centering
  \includegraphics[width=0.99\textwidth, angle=0]{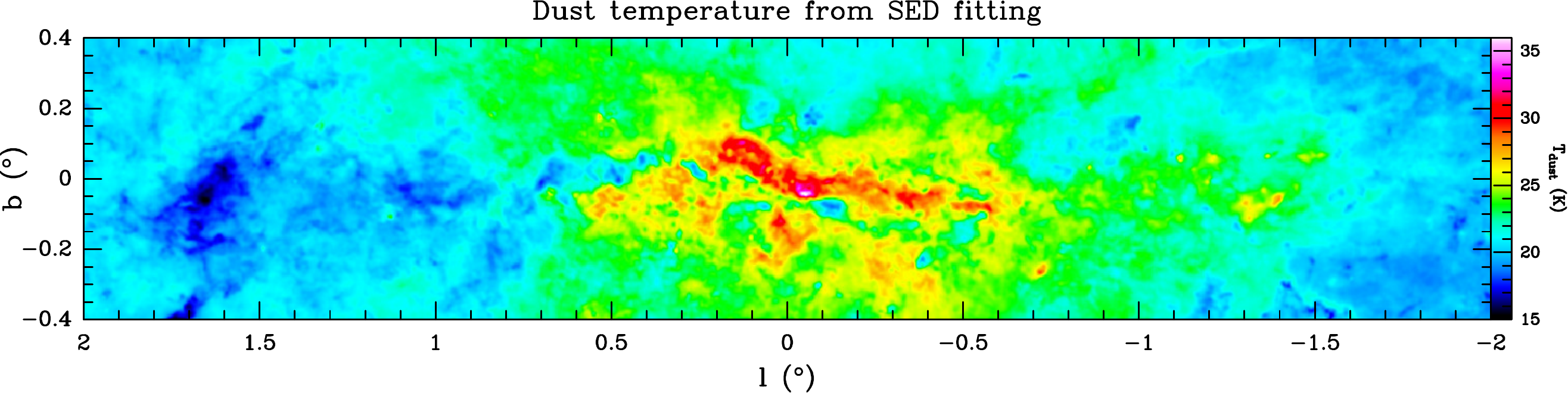}
  \caption{Dust temperature via fitting the SED of the multiwavelength observations on a pixel-by-pixel basis using the high-quality \texttt{Herschel} data whose wavelength ranges from 70 to 500\,$\mu$m.}
  \label{Fig_Tdust}
\end{figure*}

\newpage

\begin{table*}
  \caption{Parameters of identified Gaussian clumps.}
  \label{tab_clumps} \centering 
  \begin{tabular}{cccccccccc}
    \hline \hline
    Clumps & $l$, $b$                  & FWHM     & $R_{\rm eff}$ & $T_{\rm dust}$   & $I_{\rm 870{\mu}m}$ & $S_{\rm 870{\mu}m}$ & $M_{\rm H_{2}}$    & $N_{\rm H_{2}}$       & $n_{\rm H_{2}}$    \\
    No.    & $(^\circ,\,^\circ)$         & $('')$     & (pc)            & (K)                & $(\jyb)$              & (Jy)                  & $(10^{3} \Msun)$     & $(\rm 10^{23} cm^{-2})$ & $(\rm 10^4 cm^{-3})$ \\
    \hline
    1      & $ ( 0.6765 , -0.0268 ) $  & $ 29.1 $ & $ 0.68 $      & $ 23.6 \pm 2.4 $ & $ 139.82 $          & $ 331.45 $          & $ 92.2 \pm 12.8 $  & $ 28.8 \pm 4.0 $      & $ 51.9 \pm 7.2 $   \\
    2      & $ ( 0.6649 , -0.0334 ) $  & $ 34.1 $ & $ 0.80 $      & $ 24.3 \pm 2.4 $ & $ 127.47 $          & $ 405.40 $          & $ 108.2 \pm 14.9 $ & $ 24.5 \pm 3.4 $      & $ 37.7 \pm 5.2 $   \\
    3      & $ ( 0.6545 , -0.0406 ) $  & $ 42.3 $ & $ 0.99 $      & $ 23.0 \pm 2.3 $ & $ 46.17 $           & $ 266.50 $          & $ 76.7 \pm 10.8 $  & $ 11.3 \pm 1.6 $      & $ 14.0 \pm 2.0 $   \\
    4      & $ ( 0.6667 , -0.0203 ) $  & $ 42.0 $ & $ 0.98 $      & $ 21.2 \pm 2.1 $ & $ 31.21 $           & $ 217.21 $          & $ 70.1 \pm 10.1 $  & $ 10.5 \pm 1.5 $      & $ 13.1 \pm 1.9 $   \\
    5      & $ ( 0.6923 , -0.0268 ) $  & $ 71.2 $ & $ 1.66 $      & $ 19.9 \pm 2.0 $ & $ 21.92 $           & $ 325.98 $          & $ 115.1 \pm 16.9 $ & $ 6.0 \pm 0.9 $       & $ 4.4 \pm 0.7 $    \\
    6      & $ ( 0.6442 , -0.0506 ) $  & $ 51.6 $ & $ 1.20 $      & $ 20.7 \pm 2.1 $ & $ 15.65 $           & $ 130.80 $          & $ 43.7 \pm 6.3 $   & $ 4.3 \pm 0.6 $       & $ 4.4 \pm 0.6 $    \\
    7      & $ ( -0.1330 , -0.0824 ) $ & $ 58.9 $ & $ 1.37 $      & $ 18.7 \pm 1.9 $ & $ 10.82 $           & $ 130.81 $          & $ 50.9 \pm 7.7 $   & $ 3.9 \pm 0.6 $       & $ 3.5 \pm 0.5 $    \\
    8      & $ ( -0.3844 , -0.2427 ) $ & $ 32.5 $ & $ 0.76 $      & $ 21.9 \pm 2.2 $ & $ 11.09 $           & $ 32.06 $           & $ 9.9 \pm 1.4 $    & $ 2.5 \pm 0.4 $       & $ 4.0 \pm 0.6 $    \\
    9      & $ ( 0.6750 , -0.0056 ) $  & $ 68.5 $ & $ 1.60 $      & $ 18.8 \pm 1.9 $ & $ 10.59 $           & $ 141.75 $          & $ 54.7 \pm 8.2 $   & $ 3.1 \pm 0.5 $       & $ 2.4 \pm 0.4 $    \\
    10     & $ ( 0.6295 , -0.0604 ) $  & $ 69.9 $ & $ 1.63 $      & $ 19.9 \pm 2.0 $ & $ 10.26 $           & $ 142.89 $          & $ 50.5 \pm 7.4 $   & $ 2.7 \pm 0.4 $       & $ 2.1 \pm 0.3 $    \\
    11     & $ ( 0.4769 , -0.0058 ) $  & $ 56.2 $ & $ 1.31 $      & $ 18.2 \pm 1.8 $ & $ 9.39 $            & $ 85.24 $           & $ 34.5 \pm 5.2 $   & $ 2.9 \pm 0.4 $       & $ 2.7 \pm 0.4 $    \\
    12     & $ ( 0.6761 , -0.0395 ) $  & $ 27.5 $ & $ 0.64 $      & $ 25.1 \pm 2.5 $ & $ 10.38 $           & $ 50.32 $           & $ 12.9 \pm 1.8 $   & $ 4.5 \pm 0.6 $       & $ 8.6 \pm 1.2 $    \\
    13     & $ ( 0.6536 , -0.0148 ) $  & $ 41.6 $ & $ 0.97 $      & $ 19.4 \pm 1.9 $ & $ 8.48 $            & $ 56.04 $           & $ 20.7 \pm 3.1 $   & $ 3.2 \pm 0.5 $       & $ 4.0 \pm 0.6 $    \\
    14     & $ ( -0.5590 , -0.1044 ) $ & $ 41.0 $ & $ 0.95 $      & $ 21.5 \pm 2.2 $ & $ 8.01 $            & $ 40.18 $           & $ 12.7 \pm 1.8 $   & $ 2.0 \pm 0.3 $       & $ 2.6 \pm 0.4 $    \\
    15     & $ ( -0.0571 , -0.0455 ) $ & $ 64.2 $ & $ 1.49 $      & $ 36.2 \pm 3.6 $ & $ 7.86 $            & $ 94.26 $           & $ 14.9 \pm 1.9 $   & $ 1.0 \pm 0.1 $       & $ 0.8 \pm 0.1 $    \\
    16     & $ ( -0.1062 , -0.0703 ) $ & $ 74.9 $ & $ 1.74 $      & $ 19.2 \pm 1.9 $ & $ 7.74 $            & $ 125.70 $          & $ 47.1 \pm 7.0 $   & $ 2.2 \pm 0.3 $       & $ 1.6 \pm 0.2 $    \\
    17     & $ ( 0.2551 , 0.0160 ) $   & $ 70.6 $ & $ 1.64 $      & $ 19.7 \pm 2.0 $ & $ 6.93 $            & $ 154.26 $          & $ 55.6 \pm 8.2 $   & $ 2.9 \pm 0.4 $       & $ 2.2 \pm 0.3 $    \\
    18     & $ ( 0.6621 , -0.0449 ) $  & $ 22.9 $ & $ 0.53 $      & $ 24.4 \pm 2.4 $ & $ 7.29 $            & $ 11.45 $           & $ 3.0 \pm 0.4 $    & $ 1.5 \pm 0.2 $       & $ 3.5 \pm 0.5 $    \\
    19     & $ ( -0.0233 , -0.0699 ) $ & $ 89.3 $ & $ 2.08 $      & $ 23.0 \pm 2.3 $ & $ 6.32 $            & $ 179.16 $          & $ 51.5 \pm 7.2 $   & $ 1.7 \pm 0.2 $       & $ 1.0 \pm 0.1 $    \\
    20     & $ ( 0.3758 , 0.0410 ) $   & $ 31.3 $ & $ 0.73 $      & $ 22.7 \pm 2.3 $ & $ 5.95 $            & $ 15.93 $           & $ 4.7 \pm 0.7 $    & $ 1.3 \pm 0.2 $       & $ 2.1 \pm 0.3 $    \\
    21     & $ ( 0.6816 , -0.0179 ) $  & $ 24.2 $ & $ 0.56 $      & $ 19.9 \pm 2.0 $ & $ 8.17 $            & $ 15.91 $           & $ 5.6 \pm 0.8 $    & $ 2.5 \pm 0.4 $       & $ 5.5 \pm 0.8 $    \\
    22     & $ ( 1.1255 , -0.1083 ) $  & $ 38.9 $ & $ 0.91 $      & $ 24.0 \pm 2.4 $ & $ 5.32 $            & $ 24.84 $           & $ 6.7 \pm 0.9 $    & $ 1.2 \pm 0.2 $       & $ 1.6 \pm 0.2 $    \\
    23     & $ ( 0.4934 , 0.0179 ) $   & $ 61.4 $ & $ 1.43 $      & $ 19.0 \pm 1.9 $ & $ 5.19 $            & $ 59.99 $           & $ 22.7 \pm 3.4 $   & $ 1.6 \pm 0.2 $       & $ 1.4 \pm 0.2 $    \\
    ...                                                                                                                                                                                            \\
    \multicolumn{5}{l}{Others are listed only in online table.}                                                                                                                                    \\
    \hline
  \end{tabular}
\end{table*}

\section{Estimation of clump density}\label{estimate:rho}
{
  A crucial step in our analysis is to estimate the density of the clumps. Ideally,
  for a clump of a constant density, its density can be estimated using
  \begin{equation}
    \rho_{\rm 0} = \frac{m_{\rm 0}}{4/3\, \pi\, r_{\rm
          0}^3}\;.
  \end{equation}
  where $r_0$ is the clump radius and $m_0$ is the mass. However, in our
  analysis,  {due to the fact that we can only trace the distribution of gas
      in 2D, as well as our clump extraction procedure,} both $m_{\rm clump}$ and
  $r_{\rm clump}$ might have some biases. To access these effects, we have
  performed a simulation where we created a clump of a constant density in 3D,
  projected it to 2D, and computed the mass and size of the simulated clump  {by recovering it using the \texttt{GAUSSCLUMPS} algorithm}.
  Assuming that the original clump has a mass of $m_{\rm 3D}$ and size of
  $r_{\rm 3D}$, and the  {recovered} clump has a mass of $m_{\rm
        clump}$  and a size of $r_{\rm clump}$, we find
  \begin{equation}
    m_{\rm clump} = 0.76 \times m_{\rm 3D}\;,
  \end{equation}
  and
  \begin{equation}
    r_{\rm clump} = 0.72 \times r_{\rm 3D}\;.
  \end{equation}
  To accurately derive the clump density, we propose to use the equation
  \begin{equation}
    \rho_{\rm clump} = 0.12 \times m_{\rm clump}\, r_{\rm clump}^{-3}\;,
  \end{equation}
  such that
  $\rho_{\rm 3D} = \rho_{\rm clump}$ is guaranteed.
}

\bibliographystyle{aasjournal}



\end{document}